\documentclass[pre,aps,twocolumn,amsmath,amssymb,showpacs]{revtex4}
\usepackage{graphicx}
\usepackage{epsfig}
\include{graphics}
\begin{document}

\title{Diverging probability density functions 
for flat-top solitary waves}

\author{Avner Peleg$^{1}$, Yeojin Chung$^{2}$, Tom\'{a}\v{s} Dohnal$^{3}$, 
and Quan M. Nguyen$^{1}$}

\affiliation{$^{1}$Department of Mathematics, State University of New York
at Buffalo, Buffalo, New York 14260, USA}
\affiliation{$^{2}$Department of Mathematics, Southern Methodist University, 
Dallas, Texas 75275, USA}
\affiliation{$^{3}$Institute for Applied and Numerical Mathematics 2, 
Universit\"at Karlsruhe, Karlsruhe 76128, Germany}


\begin{abstract}
We investigate the statistics of flat-top solitary wave parameters in the 
presence of weak multiplicative dissipative disorder. We consider first 
propagation of solitary waves of the cubic-quintic nonlinear 
Schr\"odinger equation (CQNLSE) in the presence of disorder in 
the cubic nonlinear gain. We show by a perturbative analytic calculation and 
by Monte Carlo simulations that the probability 
density function (PDF) of the amplitude $\eta$ exhibits 
loglognormal divergence near the maximum possible amplitude $\eta_{m}$, 
a behavior that is similar to the one observed earlier for disorder in the 
linear gain [A. Peleg et al., Phys. Rev. E {\bf 72}, 027203 (2005)]. 
We relate the loglognormal divergence of the amplitude PDF to the 
super-exponential approach of $\eta$ to $\eta_{m}$ in the corresponding 
deterministic model with linear/nonlinear gain.  
Furthermore, for solitary waves of the derivative 
CQNLSE with weak disorder in the linear gain both the amplitude  
and the group velocity $\beta$ become random. We therefore study analytically 
and by Monte Carlo simulations the PDF 
of the parameter $p$, where $p=\eta/(1-\varepsilon_s\beta/2)$ and 
$\varepsilon_s$ is the self-steepening coefficient. Our analytic calculations 
and numerical simulations show that the PDF of $p$ is loglognormally 
divergent near the maximum $p$-value.
             
\end{abstract}

\maketitle

\section{Introduction}
\label{Introduction}
Flat-top solitary waves are coherent patterns, which 
exist as a result of a balance between dispersion/diffraction 
and competing nonlinearities, where the low order nonlinearity 
is ``focusing'' while the high order nonlinearity is ``defocusing''
\cite{Pushkarov96,Kivshar98,Grimshaw2001}. 
When the intensity of the field is relatively small, the low order 
nonlinearity is dominant, and consequently, the solitary waves are narrow 
and have a shape that is similar to that of conventional solitons. 
However, when the intensity increases, the high order nonlinearity 
becomes dominant and leads to the broadening of the pulse shape and 
to the generation of a typical table-top pattern.     
Flat-top solitary waves appear as solutions to 
nonlinear wave models in many areas of physics, 
including nonlinear optics \cite{Pushkarov96,Gagnon89,Akhmediev97} 
fluid dynamics \cite{Grimshaw2001,Grimshaw2002}, 
and plasma physics \cite{He92,Tribeche2008}. As a result, 
they have been the subject of intensive research efforts in recent years. 
The interest in flat-top solitary waves is further enhanced since they
are used in pattern formation theory to explain the emergence of 
fronts (kinks) from localized coherent structures such as 
solitons and solitary waves \cite{Hohenberg92,Malomed90}. 
Many of the systems in which flat-top 
solitary waves appear can be influenced by processes involving noise 
or disorder. When the disorder is strong the solitary waves are usually 
destroyed, whereas, when it is weak, the solitary waves can form and evolve. 
In the latter case one is mainly concerned with the statistics of the 
solitary wave parameters.

In this study we focus attention on an important type of disorder, 
which we call multiplicative dissipative disorder. This type of 
disorder is characterized by the following properties: 
(1) the disorder affects the amplitude of the solitary wave in first 
order of the perturbation; (2) the disorder term in the nonlinear wave 
model is multiplicative. Dissipative disorder can appear in systems 
described by nonlinear wave equations in various forms. Two of the 
most common forms are disorder in the linear gain/loss coefficient  
and disorder in the cubic nonlinear gain/loss coefficient. 
Disorder in the linear gain coefficient 
can appear in optical fiber communication systems due to randomness 
in the gain of amplifiers that are positioned along the fiber 
line to compensate for the loss \cite{Kodama83}. Moreover, such 
disorder appears in massive multichannel optical fiber transmission as 
a result of the interplay between Raman-induced energy exchange 
in pulse collisions and bit pattern randomness 
\cite{Tkach95,Ho2000,P2004,CP2005,CP2008}. 
In this case, the disorder can lead to relatively high bit-error-rate 
values and intermittent dynamics of pulse parameters 
\cite{CP2008,P2007,Cascade}. We point out that in all the studies reported 
in Refs. \cite{Kodama83,Tkach95,Ho2000,P2004,CP2005,CP2008,P2007,Cascade} 
weak disorder was considered. In addition, both linear and cubic nonlinear 
disorder in the gain can emerge in an active nonlinear 
medium due to random variations with distance in the linear/nonlinear 
gain/loss coefficient.

We consider two nonlinear wave models, 
which possess flat-top solitary wave solutions: 
the cubic-quintic nonlinear Schr\"odinger equation (CQNLSE) and  
the derivative CQNLSE (DCQNLSE). A third model, the extended Korteweg-de 
Vries (eKdV) equation, is also briefly discussed.  
The CQNLSE is a simple nonintegrable extension of the 
cubic nonlinear Schr\"odinger equation (CNLSE) 
possessing solitary wave solutions. The CQNLSE describes a variety of physical 
systems including pulse propagation in semiconductor-doped optical fibers 
\cite{Agrawal2001,Mihalache88,Gagnon89,Herrmann92,Pushkarov96,Kartashov2004,
Biswas2006}, laser-plasma interaction \cite{He92,He2004}, 
and Bose-Einstein condensates  
\cite{Newman2000,Tanatar2000,Tang2007,Kevrekidis2008}. 
Another important reason for the interest in the CQNLSE is that due 
to its nonintegrability it allows one to observe dynamical 
effects that do not exist in the CNLSE, e.g., emission of 
continuous radiation in two-soliton collisions 
\cite{Malomed86,SP2004}. Furthermore, the cubic-quintic complex 
Ginzburg-Landau equation, which is a generalization of the 
CQNLSE including dissipative terms, is known to describe even a 
wider range of physical systems, including, for example, 
convection and pattern formation in fluids 
\cite{Hohenberg92,Pomeau90,Malomed90,Deissler94,Kramer2002,Kramer2004}  
and mode-locked lasers 
\cite{Moores93,Moores95,Akhmediev97,Akhmediev2005,Kutz2006}.

The derivative CNLSE is an extension of the CNLSE that, in the 
context of nonlinear optics, takes into account the effects of 
self-steepening \cite{Agrawal2001,Tzoar81,Lisak83,Shen84}. 
For short optical pulses propagating in semiconductor-doped 
fibers both quintic nonlinearity and self-steepening are important. 
Consequently, one expects that the derivative CQNLSE, which takes 
into account both effects, would provide a more accurate  
description of the propagation in this case. 
We note that a variant of the derivative CNLSE is known to describe 
propagation of Alfv\'en waves in magnetized plasmas 
\cite{Mio76,Mjolhus76,Kaup78}. Moreover, it was recently 
shown theoretically and experimentally that a variant of the 
DCQNLSE accurately describes propagation 
of high-intensity pulses in cascaded-quadratic nonlinear media 
\cite{Wise2007}.

The eKdV equation, which is also known as the Gardner equation, 
is an integrable model that describes interfacial waves in a two-layer 
system \cite{Kakutani78,Koop81,Ablowitz81} as well as stratified 
shear flow in the ocean \cite{Lee79,Grimshaw2001,Grimshaw2002a}. 
It provides a possible explanation to observations of 
large-amplitude flat-top solitary waves in coastal zones 
\cite{Stanton98,Jeans2001,Holloway2001}.

In a previous work \cite{PDC2005} we studied the effects of weak 
disorder in the linear gain coefficient on solitary waves of the 
CQNLSE. We showed analytically (by employing an adiabatic perturbation method) 
and by Monte Carlo simulations that the probability density function (PDF) 
of the solitary wave amplitude has a loglognormal 
diverging form in the vicinity of the maximum possible amplitude. 
Since solitary waves with amplitude values close to the maximum 
possible amplitude have a table-top shape, this finding means that 
the amplitude PDF of flat-top solitary waves exhibits loglognormal 
divergence. We also conjectured that similar 
loglognormal divergence should be observed for disorder in the nonlinear 
gain. However, the full analytic form of the amplitude PDF for disorder 
in the nonlinear gain was not obtained and the conjecture was not tested by 
numerical simulations. Thus, the important question concerning the 
generality of the loglognormal divergence of the amplitude PDF 
for flat-top solitary waves was not fully answered, 
even for the CQNLSE. Furthermore, it is not clear whether 
loglognormal divergence can be observed in 
other nonlinear wave models possessing solitary wave solutions.
In the current paper we address these questions in detail. We start by 
considering propagation of solitary waves of the CQNLSE in the presence 
of weak disorder in the cubic nonlinear gain. The case of disorder in 
the cubic nonlinear gain is particularly important since 
cubic gain/loss is very common in many systems described by the 
complex Ginzburg-Landau equation (see, for example, Ref. \cite{Kramer2002} 
and references therein). We calculate the amplitude PDF 
analytically by employing an adiabatic perturbation method and validate 
its loglognormal divergence by Monte Carlo simulations. 
We then turn to study propagation 
of solitary waves of the DCQNLSE in the presence 
of weak disorder in the linear gain. In this case both the amplitude and 
the group velocity randomly vary during propagation. We therefore study 
the PDF of a new parameter, which is the ratio between the amplitude 
and a linear function of the group velocity. We find that the PDF of this 
new parameter is loglognormally divergent near the parameter's maximum value. 
We conclude by a brief discussion of the dynamic mechanism responsible 
for the loglognormal divergence of the PDFs, and of the possibility 
to observe similar statistical behavior in the eKdV model.

The material in the rest of the paper is organized as follows. 
In Sec. \ref{CQNLS},  we study propagation of solitary waves of the 
CQNLSE in the presence of disorder in the cubic nonlinear gain.
In Sec. \ref{DCQNLS}, we investigate propagation of solitary waves
of the DCQNLSE in the presence of disorder in the linear gain. Sections 
\ref{discussion} and \ref{conclusions} are reserved for discussion and 
conclusions, respectively. In Appendix \ref{G_v_g}, we describe a method 
for identifying loglognormal divergence in numerical data. Finally, 
Appendixes \ref{FPE} and \ref{Fd_Ito} are devoted to calculation 
of the PDF by employing the Fokker-Planck approach and by 
working within It\^o's interpretation.

\section{Cubic-quintic NLSE with disorder in the cubic gain/loss coefficient}
\label{CQNLS}
Consider the dynamics described by the CQNLSE with disorder 
in the cubic gain/loss coefficient:  
\begin{eqnarray}
i\partial_{z}\psi+\partial_{t}^{2}\psi+2|\psi|^2\psi -
\varepsilon_q|\psi|^4\psi=i\epsilon\xi(z)|\psi|^2\psi,
\label{div1}
\end{eqnarray}
where the disorder $\xi(z)$ is zero in average and 
short correlated in $z$:
\begin{eqnarray}
\langle\xi(z)\rangle=0,\quad
\langle\xi(z)\xi(z')\rangle=D\delta(z-z').
\label{div2}
\end{eqnarray} 
In the context of nonlinear optical waveguides $\psi$ is proportional to 
the envelope of the electric field, $z$ is the propagation distance, 
$t$ is a retarded time, $\varepsilon_{q}$ is the quintic nonlinearity 
coefficient, $0<\epsilon\ll 1$ is the cubic gain coefficient and $D$ is 
the disorder intensity. The terms $\varepsilon_q|\psi|^4\psi$ and
$i\epsilon\xi(z)|\psi|^2\psi$ describe the effects of quintic 
nonlinearity and disorder in the cubic nonlinear gain/loss coefficient, 
respectively. When $\epsilon=0$, Eq. (\ref{div1}) supports 
stable solitary wave solutions of the form \cite{Akhmediev99}: 
$\psi_{s}(t,z)=\Psi_{s}(x)\exp(i\chi)$, where
\begin{eqnarray} &&
\Psi_{s}(x) =
\frac{\sqrt{2}\eta}
{\left[(1-\eta^{2}/\eta_{m}^{2})^{1/2}\cosh(2x)+1\right]^{1/2}},
\;\;\;\;\;\;
 \label{div3}
\end{eqnarray}
$\eta_{m}\equiv(4\varepsilon_{q}/3)^{-1/2}$, 
$\chi = \alpha+\beta(t-y)+(\eta^2-\beta^2)z$, and
$x=\eta(t-y-2\beta z)$. In these relations the parameters $\eta,
\beta,\alpha,y$ are related to the amplitude, frequency, phase and
position of the solitary wave, respectively. Note that the
solitary wave solution $\psi_{s}$ exists provided that 
$\eta<\eta_{m}$.

\begin{figure}[ptb]
\epsfxsize=7.5cm  \epsffile{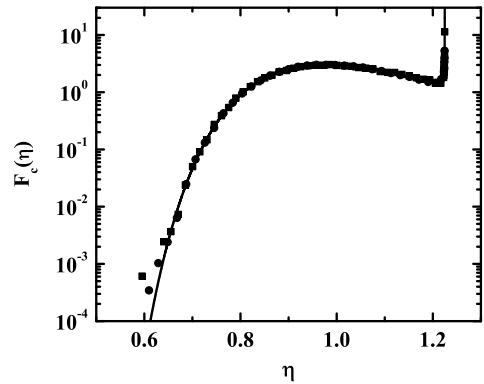}
\caption{The probability density function $F_{c}(\eta)$ at $z=10$ 
for $D=3$, $\varepsilon_{q}=0.5$, $\epsilon=0.03$ and $\eta(0)=1$.
The solid curve corresponds to the analytic result obtained by using 
Eqs. (\ref{div6}) and (\ref{div8}). The squares represent the 
result of Monte Carlo simulations with Eq. (\ref{div1}), 
while the circles stand for the result of Monte Carlo simulations 
with Eq. (\ref{div6}).}  
\label{fig1}
\end{figure}

\begin{figure}[ptb]
\epsfxsize=7.5cm  \epsffile{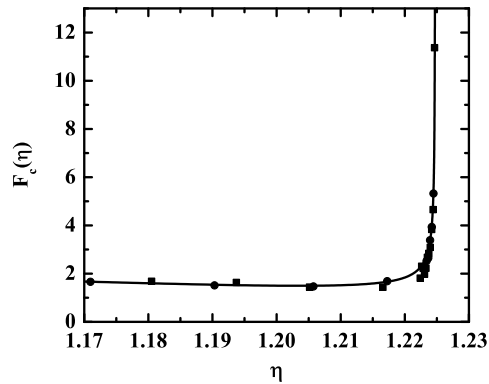}
\caption{Blowup of the data in Fig. \ref{fig1} 
in the vicinity of $\eta_{m}$ showing the divergence of $F_{c}(\eta)$.}  
\label{fig2}
\end{figure}

We study the dynamics of the solitary wave $\psi_{s}$ as described by 
Eq. (\ref{div1}). Since we are interested in flat-top solitary 
waves we focus attention on the case $\varepsilon_q>0$. 
We also assume that $4D\epsilon^{2}z\ll 1$, so that for most 
of the disorder realizations the dynamics of the solitary wave 
amplitude is not yet influenced by the $O(\epsilon^{2})$ radiation 
instability effects \cite{Hohenberg92,Stability}.
The dynamics of the parameter $\eta$ is obtained by
using energy balance considerations:
\begin{eqnarray}&&
\partial_{z}\int_{-\infty}^{\infty} dt|\psi|^{2}=
2\epsilon\xi(z)\int_{-\infty}^{\infty}dt|\psi|^{4}.
 \label{div4}
\end{eqnarray}
In order to solve Eq. (\ref{div4}) we employ the adiabatic 
perturbation method, which has been extensively used in previous 
studies of the CQNLSE, see, for example, Ref. \cite{Hohenberg92}. 
This calculation yields: 
\begin{eqnarray}&&
\!\!\!\!\!\!\!\!\!\!\!\!
\frac{d}{dz} \left[\mbox{arctanh}
\left(\frac{\eta}{\eta_{m}}\right)\right]= 
4\epsilon\eta_{m}^{2}\xi(z)
 \nonumber \\ &&
\times\left[\mbox{arctanh}
\left(\frac{\eta}{\eta_{m}}\right)-\frac{\eta}{\eta_{m}}\right].    
\label{div5}
\end{eqnarray}
Furthermore, within the framework of Stratonovich's interpretation
\cite{Stratonovich63,Gardiner2004,VanKampen2007}  
Eq. (\ref{div5}) can be transformed into 
\begin{eqnarray} &&
\!\!\!\!\!\!\!\!
\frac{d \eta}{dz}=
4\epsilon\xi(z)\left(\eta_{m}^{2}-\eta^{2}\right)
\left[\eta_{m}\mbox{arctanh}
\left(\frac{\eta}{\eta_{m}}\right)-\eta\right].    
 \label{div5a}
\end{eqnarray}  
Changing variables from $\eta$ to $v$, where  
\begin{eqnarray} &&
\!\!\!\!\!\!\!\!
v=\int\frac{d\eta}
{\left(\eta_{m}^{2}-\eta^{2}\right)
\left[\eta_{m}\mbox{arctanh}\left(\eta/\eta_{m}\right)
-\eta\right]},    
 \label{div5b}
\end{eqnarray}   
we obtain the equation $dv/dz=4\epsilon\xi(z)$, whose solution is 
\begin{eqnarray} &&
\!\!\!\!\!\!\!\!
v(z)-v(0)=\int_{\eta(0)}^{\eta(z)} \frac{d\eta}
{\left(\eta_{m}^{2}-\eta^{2}\right)
\left[\eta_{m}\mbox{arctanh}\left(\eta/\eta_{m}\right)
-\eta\right]}
 \nonumber \\ &&
=4\epsilon X(z),    
 \label{div6}
\end{eqnarray} 
where $X(z)=\int_{0}^{z}{\rm d}z'\xi(z')$. Notice that $X(z)$ can 
be regarded as a sum over many independent random variables. 
Consequently, according to the central limit theorem 
its PDF approaches a Gaussian PDF of the form
\begin{eqnarray} &&
\tilde F(X)=(2\pi Dz)^{-1/2}\exp\left[-X^{2}/(2Dz)\right].     
 \label{div7}
\end{eqnarray}     
Equation (\ref{div6}) defines a monotonously increasing function 
$X=q(\eta)$ on $0\le \eta < \eta_{m}$.  
Changing variable from $X$ to $\eta$ while employing 
Eqs. (\ref{div6}) and (\ref{div7}) we obtain that the PDF of 
$\eta$ is given by 
\begin{eqnarray} &&
F_{c}(\eta)=\frac{(32\pi D\epsilon^{2}z)^{-1/2}
\exp\left[-q^{2}(\eta)/(2Dz)\right]}
{(\eta_{m}^{2}-\eta^{2})
\left[\eta_{m}\mbox{arctanh}\left(\eta/\eta_{m}\right)-\eta\right]}     
 \label{div8}
\end{eqnarray}       
for $0<\eta<\eta_{m}$ and $F_{c}(\eta)=0$ elsewhere.
To calculate the value of $F_{c}(\eta)$ for a given $\eta$ one numerically 
solves Eq. (\ref{div6}) for $X=q(\eta)$ and then substitutes the result  
into Eq. (\ref{div8}). The graph of $F_{c}(\eta)$ obtained 
by this calculation is shown in Figs. \ref{fig1} and \ref{fig2}.


As can be seen from Figs. \ref{fig1} and  \ref{fig2}, $F_{c}(\eta)$ 
diverges in the vicinity of $\eta_{m}$. In order to characterize 
the divergence we obtain an approximate analytic solution of 
Eq. (\ref{div6}) for $\eta$ near $\eta_{m}$. We first write the 
integral in Eq. (\ref{div6}) as a sum of two integrals:
 \begin{eqnarray}&&
\!\!\!\!\!\!\!\!
4\epsilon X(z)=\int_{\eta(0)}^{\eta(\tilde z)} \frac{d\eta}
{\left(\eta_{m}^{2}-\eta^{2}\right)
\left[\eta_{m}\mbox{arctanh}\left(\eta/\eta_{m}\right)
-\eta\right]}
 \nonumber \\&&
+\int_{\eta(\tilde z)}^{\eta(z)} \frac{d\eta}
{\left(\eta_{m}^{2}-\eta^{2}\right)
\left[\eta_{m}\mbox{arctanh}\left(\eta/\eta_{m}\right)
-\eta\right]},    
 \label{div8a}
\end{eqnarray} 
where $\eta(\tilde z)$ is a constant satisfying 
$\eta(\tilde z)<\eta(z)<\eta_{m}$, and both $\eta(\tilde z)$ and 
$\eta(z)$ are close to $\eta_{m}$. Since both limits of the first integral 
on the right hand side of Eq. (\ref{div8a}) are constants this integral 
is a constant that we denote by $c_{1}$.  
We also denote $\delta\eta=\eta_{m}-\eta$ and notice 
that $0<\delta\eta(z)/\eta_{m}<\delta\eta(\tilde z)/\eta_{m}\ll 1$.   
We can therefore expand the integrand in the second integral on the right 
hand side of Eq. (\ref{div8a}) about $\delta\eta=0$, keeping terms 
up to order $\delta\eta$ in the denominator. This calculation yields 
\begin{eqnarray}&&
\!\!\!\!\!4\epsilon X(z)\simeq c_{1}+\frac{1}{\eta_{m}^{2}}
\int_{\delta\eta(\tilde z)}^{\delta\eta(z)} \frac{d \delta\eta}
{\delta\eta\ln\left(\frac{e^{2}\delta\eta}{2\eta_{m}}\right)}.
 \label{div9}
\end{eqnarray} 
Integrating over $\delta\eta$ we arrive at
\begin{eqnarray}&&
X(z)\simeq
\ln\left\{-\ln\left[\frac{e^{2}\delta\eta(z)}{2\eta_{m}}\right]
/\tilde c\right\}/(4\epsilon\eta_{m}^{2}),
 \label{div10}
\end{eqnarray} 
where $\tilde c$ is another constant. Since the normally distributed 
random variable $X(z)$ is related to $\delta\eta(z)$ via a double logarithm, 
we say that $\delta\eta(z)$ is loglognormally distributed.  
Using Eqs. (\ref{div7}) and (\ref{div10}),   
and changing variables from $X$ to $\delta\eta$ we obtain
\begin{eqnarray}&&
\!\!\!\!\!\left.F_{c}(\eta)\right|_{\eta\lesssim
\eta_{m}}\!\!\!\!\simeq\! 
\left\{(32\pi D\epsilon^{2}z)^{1/2}\eta_{m}^{2}\delta\eta
\left|\ln\left[\frac{e^{2}\delta\eta}{2\eta_{m}}\right]\right|
\right\}^{-1}
 \nonumber \\ &&
\!\!\!\!\!\times\!\exp\left\{\!-\frac{\ln^{2}\!\left[
-\ln\left[(e^{2}\delta\eta)/(2\eta_{m})\right]/{\tilde c}\right]} 
{32D\epsilon^{2}\eta_{m}^{4}z}\right\},
 \label{div11}
\end{eqnarray}
from which it follows that the divergence of $F_{c}(\eta)$ near $\eta_{m}$  
is loglognormal.

To check the analytic predictions given by Eqs. (\ref{div8}) and 
(\ref{div11}) we performed Monte Carlo simulations with Eq. 
(\ref{div1}) with $1.09\times 10^{5}$ disorder realizations. 
We considered the parameter values $D=3$, $\varepsilon_{q}=0.5$ 
(corresponding to $\eta_{m}\simeq 1.22474$) and $\epsilon=0.03$. 
The initial condition was taken in the form 
of the solitary wave $\psi_{s}$ with 
$\eta(0)=1$, $\beta(0)=0$, $y(0)=0$, and $\alpha(0)=0$.       
We carried out the simulations up to a final distance $z_{f}=10$, 
for which the disorder strength is $4D\epsilon^{2}z_{f}=0.3$.
Equation (\ref{div1}) was integrated by employing a split-step
method that is of fourth order with respect 
to the $z$-step $dz$ \cite{Yoshida90}. 
The linear part $i\partial_{z}\psi = -\partial_{t}^{2}\psi$ 
was advanced efficiently via an evaluation of the operator 
exponential in Fourier space and the nonlinear part  
$i\partial_{z}\psi = \varepsilon_q |\psi|^4 \psi + 
(i\epsilon \xi(z)-2)|\psi|^2\psi$ 
was advanced  via a fourth order Runge-Kutta scheme. 
To overcome numerical errors resulting from radiation emission and the use of 
periodic boundary conditions we applied the method of 
artificial damping in the vicinity of the boundaries of the computational 
domain. (See Refs. \cite{SP2004,PDC2005,Kuznetsov95} for other examples 
where the same method was successfully used). The size of 
the domain was taken to be $-L\leq t \leq L$ with $L =16\pi$ so that 
the absorbing layers do not affect the dynamics of the solitary waves.
The $t$-step and $z$-step were taken as $\Delta t =0.01$ 
and $\Delta z =0.001$, respectively.

Figure \ref{fig1} shows the $\eta$-PDF obtained by the simulations 
as well as the analytic prediction obtained with Eqs. 
(\ref{div6}) and (\ref{div8}) and the PDF obtained by Monte Carlo 
simulations with Eq. (\ref{div6}). Figure \ref{fig2} shows a blowup 
of the same data in the neighborhood of $\eta_{m}$. The good agreement 
between the three results strongly indicates that the divergence is 
indeed loglognormal. In Fig. \ref{fig3} we present a more 
sensitive analysis of this divergence that is based on the 
procedure described in Appendix \ref{G_v_g} for detecting 
loglognormal divergence in numerical data. Following this procedure 
we plot $G_{c}(\delta\eta)$ versus $g_{c}(\delta\eta)$, where
$G_{c}(\delta\eta)$ and $g_{c}(\delta\eta)$ are defined by Eqs. 
(\ref{gvg4}) and (\ref{gvg5}), respectively.   
It is seen that the data obtained by numerical simulations 
with Eq. (\ref{div1}) lies on a straight line with a slope 0.97, 
which is very close to the theoretically predicted value of 1. 
Therefore, this analysis provides further support 
in favor of the loglognormal divergence of $F_{c}(\eta)$. 
Combining this observation with the 
result of Ref. \cite {PDC2005},  
we conclude that both disorder in the linear gain/loss 
coefficient and disorder in the cubic nonlinear gain/loss coefficient 
lead to the same type of divergence of the $\eta$-PDF.   
\begin{figure}[ptb]
\epsfxsize=7.5cm  \epsffile{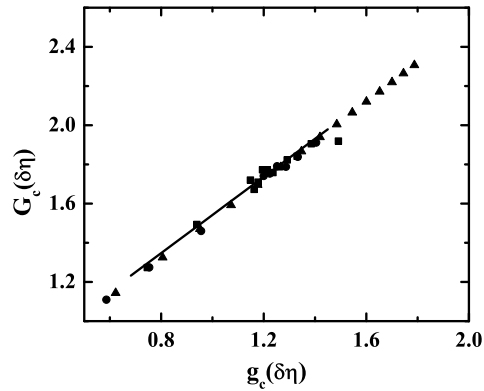}
\caption{$G_{c}(\delta \eta)$ vs $g_{c}(\delta \eta)$ 
for the same parameters  considered in Figs. \ref{fig1} and \ref{fig2}. 
The triangles represent the analytic prediction of 
Eqs. (\ref{div6}) and (\ref{div8}), the squares stand for 
the result of numerical simulations with Eq. (\ref{div1}) 
and the circles correspond to the result obtained by Monte 
Carlo simulations with Eq. (\ref{div6}). The solid line is 
a linear fit of the squares with a slope of 0.97.}  
\label{fig3}
\end{figure}

\section{Multiplicative-dissipative disorder in the derivative CQNLSE} 
\label{DCQNLS}
The discussion in section \ref{CQNLS} indicates that the 
loglognormal divergence of the amplitude PDF is quite general 
for solitary waves of the CQNLS model. We now show that 
similar statistical behavior is exhibited by solitary waves 
of a second nonlinear wave model. We consider the derivative 
CQNLSE (DCQNLSE) with weak disorder in the 
linear gain/loss coefficient
\begin{eqnarray} &&
i\partial_{z}\psi+\partial_{t}^{2}\psi+2|\psi|^2\psi -
\varepsilon_q|\psi|^4\psi+
i\varepsilon_s\partial_{t}\left(|\psi|^2\psi\right)
 \nonumber \\ &&
=i\epsilon\xi(z)\psi,
\label{SSP1}
\end{eqnarray} 
where $\xi(z)$ satisfies Eq. (\ref{div2}). In the context of nonlinear 
optics $\varepsilon_s$ is the self-steepening coefficient and 
$i\varepsilon_s\partial_{t}\left(|\psi|^2\psi\right)$ describes 
the self-steepening effect. In the absence of the perturbation term 
$i\epsilon\xi(z)\psi$, Eq. (\ref{SSP1}) possesses solitary wave solutions 
of the form     
$\psi_{s}(t,z)=\Psi_{s}(x)\exp(i\chi)$, where
\begin{eqnarray} &&
\Psi_{s}(x) =
\frac{(2-\varepsilon_{s}\beta)^{1/2}p}
{\left[(1-p^{2}/p_{m}^{2})^{1/2}\cosh(2x)+1\right]^{1/2}},
\;\;\;\;\;\;
 \label{SSP2}
\end{eqnarray}
$p=\eta/(1-\varepsilon_s\beta/2)$, 
$p_{m}=[4(\varepsilon_{q}-3\varepsilon_{s}^{2}/16)/3]^{-1/2}$,    
$\chi = \alpha+\beta(t-y)+(\eta^2-\beta^2)z+g(x)$, and 
$x=\eta(t-y-2\beta z)$. In addition, the chirp $g(x)$ is given by
\begin{eqnarray} &&
g(x)= -\frac{3}{\sqrt{4}}\varepsilon_{s}p_{m}
\mbox{arctanh}\left[B_{1}\tanh(x)\right],
 \label{SSP3}
\end{eqnarray}
where the coefficient $B_{1}$ is 
\begin{eqnarray} &&
B_{1}=\left[\frac{1-(1-p^{2}/p_{m}^{2})^{1/2}}
{1+(1-p^{2}/p_{m}^{2})^{1/2}}\right]^{1/2}.
 \label{SSP5}
\end{eqnarray} 
These solitary wave solutions exist provided that 
$\epsilon_{q}>3\epsilon_{s}^{2}/16$ and $p<p_{m}$.  
Notice that the solitary waves of the DCQNLSE are chirped 
and are thus fundamentally different from the solitary waves of 
the CQNLSE.

In the presence of disorder in the linear gain both $\eta$ 
and $\beta$ randomly vary along the propagation. Thus, the dynamics 
is different from that observed in the CQNLSE case, where only 
$\eta$ varies as a result of the disorder. 
Energy-balance considerations lead to an equation of the form 
\begin{eqnarray}&&
\partial_{z}\int_{-\infty}^{\infty} dt|\psi|^{2}=
2\epsilon\xi(z)\int_{-\infty}^{\infty}dt|\psi|^{2}.
 \label{SSP6}
\end{eqnarray}
In the first order adiabatic perturbation procedure we 
replace $\psi(t,z)$ with $\psi_{s}(t,z)$ in Eq. (\ref{SSP6})  
and obtain 
\begin{eqnarray}&&
\!\!\!\!\!\!\!\!\!\!\!\!
\frac{d}{dz} \left[\mbox{arctanh}
\left(\frac{p}{p_{m}}\right)\right]= 
2\epsilon\xi(z)
\mbox{arctanh}\left(\frac{p}{p_{m}}\right). 
\label{SSP7}
\end{eqnarray}  
Denoting $\rho_{d}(z)=\mbox{arctanh}\left[p(z)/p_{m}\right]$, 
we observe that $\rho_{d}$ satisfies the stochastic equation 
\begin{eqnarray}&&
d\rho_{d}/dz=2\epsilon\xi(z)\rho_{d}, 
\label{Ito1}
\end{eqnarray}  
whose solution in Stratonovich's interpretation is: 
\begin{eqnarray}&&
\rho_{d}(z)=\rho_{d}(0)\exp\left[2\epsilon X(z)\right]. 
\label{SSP8}
\end{eqnarray}  
Therefore, the PDF of $\rho_{d}(z)$ is lognormal. Changing variables from 
$\rho_{d}(z)$ to $p(z)$ we obtain that the PDF of $p$ is given by 
\begin{eqnarray}
\!F_{d}(p)\!\!=\!\!\frac{\exp\left\{\!-\ln^{2}\!\left[
\mbox{arctanh}\left(p/p_{m}\right)/\rho_{d}(0)\right]/
(8D\epsilon^{2}z)\right\}} {(8\pi
D\epsilon^{2}z)^{1/2}p_{m}\left(1\!-\!p^{2}/p_{m}^{2}\right)
\mbox{arctanh}\left(p/p_{m}\right)},
 \nonumber \\ &&
 \label{SSP9}
\end{eqnarray} 
for $0<p<p_{m}$ and $F_{d}(p)=0$ elsewhere.  
In Appendix \ref{FPE} we obtain the same expression for 
$F_{d}(p)$ by using the Fokker-Planck approach. The PDF 
$F_{d}(p)$ has exactly the same form as the PDF of $\eta$ in 
systems described by the CQNLSE with disorder in the linear 
gain/loss coefficient. 
(Compare Eq. (\ref{SSP9}) with Eq. (6) in Ref. \cite{PDC2005}). 
In particular, $F_{d}(p)$ exhibits loglognormal divergence 
in the vicinity of $p_{m}$: 
\begin{eqnarray}&&
\!\!\!\!\left.F_{d}(p)\right|_{p\lesssim p_{m}}\!
\simeq\!\! \left\{(8\pi D\epsilon^{2}z)^{1/2}\delta p
\left|\ln\left[\delta p/(2p_{m})\right]\right|\right\}^{-1}
 \nonumber \\ &&
\!\!\!\!\!\!\!\!\!\!\!\times\exp\left\{\!-\ln^{2}\!\left[
-\ln\left[\delta p/(2p_{m})\right]/(2\rho_{d}(0))\right]/
(8D\epsilon^{2}z)\right\}\!,
 \label{SSP10}
\end{eqnarray}
where $\delta p=p_{m}-p$ and $0<\delta p/p_{m}\ll 1$. 
In Appendix \ref{Fd_Ito} we calculate $F_{d}(p)$ 
by applying It\^o's interpretation \cite{Gardiner2004,VanKampen2007} 
to the stochastic equation (\ref{Ito1}) satisfied by $\rho_{d}$. 
We show that in this case as well $F_{d}(p)$ 
exhibits loglognormal divergence near $p_{m}$.

To validate our theoretical predictions we performed Monte Carlo 
simulations with Eq. (\ref{SSP1}) with about $7.8\times 10^{4}$ disorder 
realizations. We used an initial condition in the form of the solitary 
wave $\psi_{s}$ with $\eta(0)=1$, $\beta(0)=0$, $y(0)=0$, and $\alpha(0)=0$ 
and considered the parameter values $D=3$, $\varepsilon_{q}=0.7$, 
$\varepsilon_{s}=0.8$, and $\epsilon=0.05$. 
For these values, $p(0)=1$ and $p_{m}\simeq 1.13715$. 
The simulations were carried out 
up to a final distance $z_{f}=11$, for which the disorder strength is  
$4D\epsilon^{2}z_{f}=0.33$. Equation (\ref{SSP1}) was integrated 
by employing the split-step method with periodic boundary conditions and 
with the same numerical scheme as described in Sec. \ref{CQNLS}. 
The size of the computational domain was taken
to be $-100\leq t \leq 100$, and the $t$-step and
$z$-step were taken as $\Delta t =0.01$ and $\Delta z =0.001$, 
respectively.

\begin{figure}[ptb]
\epsfxsize=7.5cm  \epsffile{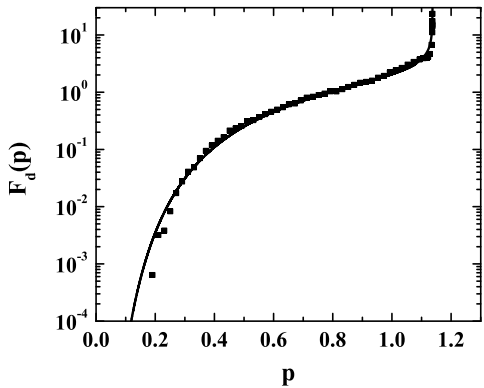}
\caption{The probability density function of $p$ 
$F_{d}(p)$ at z=10 for $D=3$, $\varepsilon_{q}=0.7$, $\varepsilon_{s}=0.8$, 
$\epsilon=0.05$ and  $p(0)=1$.  The squares represent the 
result of Monte Carlo simulations with Eq. (\ref{SSP1}), 
while the solid curve corresponds to the analytic result given by 
Eq. (\ref{SSP9}).}  
\label{SSPfig1}
\end{figure}

\begin{figure}[ptb]
\epsfxsize=7.5cm  \epsffile{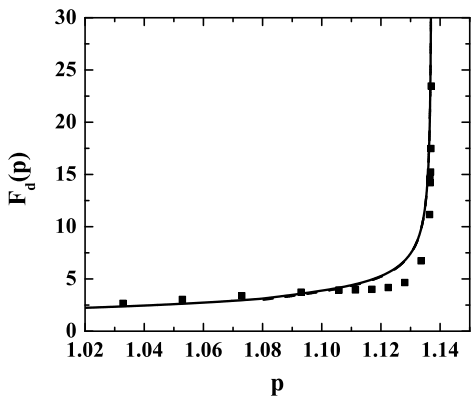}
\caption{Blowup of the data shown in Fig. \ref{SSPfig1} 
in the vicinity of $p_{m}$. The dashed line corresponds to the 
asymptotic loglognormal PDF given by Eq. (\ref{SSP10}).}   
\label{SSPfig2}
\end{figure}

\begin{figure}[ptb]
\epsfxsize=7.5cm  \epsffile{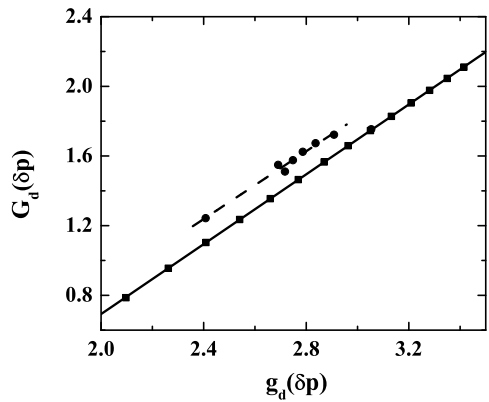}
\caption{$G_{d}(\delta p)$ vs $g_{d}(\delta p)$ for the same parameters  
considered in Figs. \ref{SSPfig1} and \ref{SSPfig2}. 
The squares represent the analytic result obtained with Eq. (\ref{SSP9}),
while the circles stand for the numerical result. The solid and dashed lines 
are linear fits with slopes 1.00 and 0.97, respectively.}   
\label{SSPfig3}
\end{figure}

The PDF $F_{d}(p)$ at $z=10$ obtained in the numerical simulations is 
shown in Fig. \ref{SSPfig1} together with the theoretical prediction. 
The numerically obtained PDF clearly exhibits divergence in the 
vicinity of $p_{m}$ and the overall agreement between theory and simulation 
is good. To further check the behavior of the PDF in the vicinity of 
$p_{m}$ Fig. \ref{SSPfig2} shows a blowup of the data in the region 
$p\lesssim p_{m}$. Reasonable agreement between theory and simulations 
is observed. We attribute the differences between the curves to 
the difficulties in obtaining an accurate measurement of the group 
velocity from the numerical data. 
As a further test for the asymptotic behavior of $F_{d}(p)$ near 
$p_{m}$ we employ the procedure for detecting loglognormal divergence 
that is outlined in Appendix \ref{G_v_g}.
Following this procedure we present the graph of $G_{d}(\delta p)$ 
versus $g_{d}(\delta p)$ in Fig. \ref{SSPfig3}, where 
$G_{d}(\delta p)=\{-\ln[(8\pi D\epsilon^{2}z)^{1/2}
\delta p|\ln[\delta p/(2p_{m})]|F_{d}(p)]\}^{1/2}$ 
and $g_{d}(\delta p)=(8D\epsilon^{2}z)^{-1/2}
\ln\{-\ln[\delta p/(2p_{m})]\}$. 
We observe that the numerically obtained 
curve lies on a straight line with a slope 0.97, which is very close to the 
theoretically predicted value of 1. We therefore conclude that the 
numerically obtained PDF of $p$ does exhibit loglognormal divergence. 

\section{Discussion} 
\label{discussion}
Here we discuss the underlying reason for  
the loglognormal divergence of the $\eta$-PDF due to disorder 
in the linear/nonlinear gain/loss. In addition, we briefly 
discuss a KdV-type of model where the loglognormal 
divergence of soliton parameters can potentially be observed.

\subsection{Loglognormal divergence of $F(\eta)$ and 
super-exponential decay of $\delta\eta$ to 0}

Consider, for example, the CQNLSE with {\it deterministic} linear gain: 
\begin{eqnarray} &&
i\partial_{z}\psi+\partial_{t}^{2}\psi+2|\psi|^2\psi -
\varepsilon_q|\psi|^4\psi=i\epsilon\psi.
\label{dis1}
\end{eqnarray}
Using the adiabatic perturbation method we obtain the following 
equation for the dynamics of $\eta$: 
\begin{eqnarray} &&
\frac{d\eta}{dz}=
2\epsilon\eta_{m}\left(1-\frac{\eta^{2}}{\eta_{m}^{2}}\right)
\mbox{arctanh}\left(\frac{\eta}{\eta_{m}}\right).
 \label{dis2}
\end{eqnarray} 
Even though Eq. (\ref{dis2}) can be solved analytically, it is 
instructive to consider its asymptotic approximation for 
$\eta\lesssim\eta_{m}$. Denoting $\delta\eta=\eta_{m}-\eta$ and  
expanding both sides of Eq. (\ref{dis2}) about $\eta_{m}$ 
while keeping terms up to $O(\delta\eta)$ we obtain 
\begin{eqnarray} &&
\frac{d\delta\eta}{dz}\simeq 
2\epsilon\delta\eta
\ln\left(\frac{\delta\eta}{2\eta_{m}}\right).
 \label{dis3}
\end{eqnarray} 
Integrating Eq. (\ref{dis3}) over $z$ we arrive at
\begin{eqnarray} &&
\delta\eta(z)\simeq 
2\eta_{m}\exp\left[-\tilde C(0) e^{2\epsilon z}\right],
 \label{dis4}
\end{eqnarray}    
where $\tilde C(0)=-\ln[\delta\eta(0)/(2\eta_{m})]>0$. 
We therefore observe that in this case $\delta\eta$ decays to 0 
super-exponentially with increasing $z$. 
It is this super-exponential approach of $\eta$ 
to $\eta_{m}$ that leads to the loglognormal divergence of $F(\eta)$. 
Indeed, for the CQNLSE with disorder in the linear gain/loss 
coefficient one obtains a similar equation for $\delta\eta(z)$ with 
$z$ replaced by $X(z)$ on the right hand side. As a result, $X(z)$ is 
related to $\delta\eta(z)$ via 
\begin{eqnarray}&&
X(z)\simeq\ln\left[-\ln\left(\frac{\delta\eta}{2\eta_{m}}\right)
/\tilde C(0)\right]/(2\epsilon),
 \label{dis5}
\end{eqnarray}   
which describes loglognormal divergence of the 
$\eta$-PDF near $\eta_{m}$. A similar result holds for the CQNLSE with 
disorder in the cubic nonlinear gain/loss coefficient 
[see Eq. (\ref{div10})].

\subsection{The extended Korteweg-de Vries equation}
A different type of nonlinear wave equation that possesses 
flat-top solitary wave solutions is the following extended 
Korteweg-de Vries (eKdV) equation 
\cite{Ablowitz81,Grimshaw2001,Grimshaw2002}:
\begin{eqnarray}
\partial_{t}u+6u(1-u)\partial_{z}u+\partial_{z}^{3}u=0.
\label{KdV1}
\end{eqnarray}  
Note that Eq. (\ref{KdV1}) is integrable \cite{Ablowitz81,Grimshaw2002}. 
In the context of interfacial waves in two-layer systems and 
stratified shear flow in the ocean $u$ represents the vertical 
displacement, $z$ is the horizontal coordinate, and $t$ is time. 
The solitary wave solutions of Eq. (\ref{KdV1}) 
take the form \cite{Grimshaw2002} 
\begin{eqnarray} &&
u_{s}(z,t) =
\frac{4\kappa^{2}}
{(1-\kappa^{2}/\kappa_{m}^{2})^{1/2}\cosh(2x)+1},
\;\;\;\;\;\;
 \label{KdV2}
\end{eqnarray}
where $x=\kappa(z-4\kappa^{2}t)$, $\kappa_{m}=1/2$, and   
the parameter $\kappa$ characterizes the soliton amplitude and 
group velocity.

Since perturbative linear gain and loss terms are quite common in 
KdV models of wave motion in the ocean \cite{Newell85,Holloway2001},  
it is interesting to study the situation 
where randomness is present. We therefore 
consider the following perturbed eKdV equation: 
\begin{eqnarray}
\partial_{t}u+6u(1-u)\partial_{z}u+\partial_{z}^{3}u=
\epsilon\xi(t)u,
\label{KdV3}
\end{eqnarray}        
where 
\begin{eqnarray}
\langle\xi(t)\rangle=0,\quad
\langle\xi(t)\xi(t')\rangle=D\delta(t-t').
\label{KdV4}
\end{eqnarray} 
Employing mass-balance considerations we obtain that the dynamics of 
$\kappa$ is described by an equation similar to Eq. (\ref{SSP7}). 
Based on this observation one would expect the $\kappa$-PDF to exhibit 
loglognormal divergence in the vicinity of $\kappa_{m}$. 
It should be pointed out, however, that for models 
of the KdV type radiative effects are more significant \cite{Newell85}, 
and as a result, a perturbative calculation that takes these effects 
into account is required. We therefore defer the full analysis of this 
case to a future publication.

\section{Conclusions}
\label{conclusions}
We studied the statistics of flat-top solitary wave parameters  
in the presence of weak dissipative multiplicative disorder. 
We started by considering propagation of 
solitary waves of the cubic-quintic nonlinear 
Schr\"odinger equation (CQNLSE) in the presence of disorder in the cubic 
nonlinear gain/loss. We found that the amplitude PDF  
exhibits loglognormal divergence in the vicinity of the maximum possible 
amplitude $\eta_{m}$. Since solitary waves with $\eta$ values near 
$\eta_{m}$ have a typical table-top shape we conclude that 
the amplitude PDF of flat-top solitary waves is 
loglognormally divergent.  
This finding combined with similar findings in Ref. \cite{PDC2005} 
for the case of disorder in the linear gain/loss coefficient indicates 
that loglognormal divergence is quite ubiquitous for flat-top solitary waves 
of the CQNLSE in the presence of weak multiplicative dissipative disorder.  
Furthermore, we showed that 
this divergence can be explained by the super-exponential approach of 
$\eta$ to $\eta_{m}$ in the corresponding deterministic model 
with weak linear/nonlinear gain.

Next we considered propagation of solitary waves in the presence of 
weak disorder in the linear gain/loss in systems described by the 
derivative cubic-quintic nonlinear Schr\"odinger equation (DCQNLSE).  
The solitary waves of the DCQNLSE are chirped and are thus 
fundamentally different from the solitary waves of the CQNLSE. 
As a result of the chirp, in the presence of disorder in the linear 
gain/loss both the amplitude $\eta$ and the group velocity $\beta$ 
vary randomly along the propagation. We therefore studied the PDF of the 
parameter $p$, where $p=\eta/(1-\varepsilon_s\beta/2)$ and $\varepsilon_s$ 
is the self-steepening coefficient. We found that $F_{d}(p)$ is  
loglognormally divergent when $p$ is near its maximum value $p_{m}$, 
i.e., the $p$-PDF of the corresponding flat-top solitary waves exhibits 
loglognormal divergence. Moreover, we showed that the same divergence 
is observed in both Stratonovich's interpretation and It\^o's 
interpretation of the linear stochastic perturbation term, 
thus illustrating another feature of the statistics that appears to be 
quite general.

\appendix
\section{The $G$ vs $g$ method}
\label{G_v_g}
Here we give the details behind the $G$ vs $g$ method that is used 
to analyze the divergence of the PDFs of $\eta$ and $p$ 
near their maximum possible values. 
As a specific example we consider the case of the CQNLSE in the 
presence of disorder in the cubic gain/loss coefficient. In Sec. \ref{CQNLS} 
we obtained the following approximate analytic expression for 
$F_{c}(\eta)$ near $\eta_{m}$: 
\begin{eqnarray}&&
\!\!\!\!\!\left.F_{c}(\eta)\right|_{\eta\lesssim
\eta_{m}}\!\!\!\!\simeq\! 
\left\{(32\pi D\epsilon^{2}z)^{1/2}\eta_{m}^{2}\delta\eta
\left|\ln\left[\frac{e^{2}\delta\eta}{2\eta_{m}}\right]\right|
\right\}^{-1}
 \nonumber \\ &&
\!\!\!\!\!\times\!\exp\left\{\!-\frac{\ln^{2}\!\left[
-\ln\left[(e^{2}\delta\eta)/(2\eta_{m})\right]/{\tilde c}\right]} 
{32D\epsilon^{2}\eta_{m}^{4}z}\right\}.
 \label{gvg1}
\end{eqnarray}
The problem we want to address is how to verify that the PDF obtained in 
the simulations satisfies the loglognormal divergence described by Eq. 
(\ref{gvg1}). In particular, we need a method that will allow us to ignore 
the coefficient $\tilde c$, which cannot be found from the numerical data. 
Furthermore, we would like to find a mapping that ``stretches'' the small 
$\eta$-neighborhood of $\eta_{m}$ into a wider interval. To address these 
issues we rewrite Eq. (\ref{gvg1}) in the form 
\begin{eqnarray}&&
\!\!\!\!\!
\ln\left[(32\pi D\epsilon^{2}z)^{1/2}\eta_{m}^{2}\delta\eta
\left|\ln\left(\frac{e^{2}\delta\eta}{2\eta_{m}}\right)\right|
F_{c}(\eta)\right]\!\simeq\! 
 \nonumber \\ &&
-\frac{\ln^{2}\!\left\{
-\ln\left[(e^{2}\delta\eta)/(2\eta_{m})\right]/{\tilde c}\right\}}
{32D\epsilon^{2}\eta_{m}^{4}z},
 \label{gvg2}
\end{eqnarray} 
where it is understood that $F_{c}(\eta)$ is calculated near $\eta_{m}$.  
Multiplying by -1 and taking the square root we arrive at
\begin{eqnarray}&&
\!\!\!\!\!
\left\{-\ln\left[(32\pi D\epsilon^{2}z)^{1/2}\eta_{m}^{2}\delta\eta
\left|\ln\left(\frac{e^{2}\delta\eta}{2\eta_{m}}\right)\right|
F_{c}(\eta)\right]\right\}^{1/2}\!\simeq\! 
 \nonumber \\ &&
\frac{\ln\!\left\{-\ln\left[(e^{2}\delta\eta)/(2\eta_{m})\right]\right\}
-\ln({\tilde c})}
{\left(32D\epsilon^{2}\eta_{m}^{4}z\right)^{1/2}}.
 \label{gvg3}
\end{eqnarray} 
We now define $G_{c}(\delta\eta)$ and $g_{c}(\delta\eta)$ as 
\begin{eqnarray}&&
G_{c}(\delta\eta)=
 \nonumber \\&&
\!\!\!\!\!\!\!\!\!\!\!\!\!\!\!\!\!\!\!\!
\left\{-\ln\left[(32\pi D\epsilon^{2}z)^{1/2}\eta_{m}^{2}\delta\eta
\left|\ln\left(\frac{e^{2}\delta\eta}{2\eta_{m}}\right)\right|
F_{c}(\eta)\right]\right\}^{1/2}\!\!\!\!\!\!\!\!\!,
 \label{gvg4}
\end{eqnarray} 
and 
\begin{eqnarray}&&
\!\!\!\!\!\!\!\!\!\!
g_{c}(\delta\eta)=
\frac{\ln\!\left\{-\ln\left[(e^{2}\delta\eta)/(2\eta_{m})\right]\right\}}
{\left(32D\epsilon^{2}\eta_{m}^{4}z\right)^{1/2}}.
 \label{gvg5}
\end{eqnarray} 
Using these definitions we observe that when the statistics of $\eta$ 
is described by Eq. (\ref{gvg1}) the graph of $G$ vs $g$ is a straight 
line with a slope 1, independent of the value of $\tilde c$. A similar 
conclusion (with slightly different expressions for $G$ and $g$) 
holds in the case of DCQNLSE with disorder in the linear 
gain/loss coefficient.

\section{Fokker-Planck approach for calculation of the PDFs}
\label{FPE}
In this Appendix we demonstrate that the expressions for the PDF of the 
solitary wave parameters that were obtained in sections \ref{CQNLS} 
and \ref{DCQNLS} by solving the Langevin equation can also 
be obtained within the framework of the Fokker-Planck approach. 
As a specific example we consider the DCQNLSE with disorder in the 
linear gain/loss coefficient and work with Stratonovich's 
interpretation of Eq. (\ref{Ito1}). 
The corresponding Fokker-Planck equation for the PDF 
of $\rho_{d}$, $H(\rho_{d},z)$, is \cite{Gardiner2004,VanKampen2007}:     
\begin{eqnarray}&&
\partial_{z}H=2D\epsilon^{2}\partial_{\rho_{d}}
\left[\rho_{d}\partial_{\rho_{d}}\left(\rho_{d}H\right)\right]. 
\label{FPE1}
\end{eqnarray}       
Changing variable to $w=\ln\rho_{d}$ we arrive at 
\begin{eqnarray}&&
\partial_{z}\tilde H=2D\epsilon^{2}\partial_{w}^{2}\tilde H,
\label{FPE2}
\end{eqnarray} 
where 
\begin{eqnarray}&&
\tilde H(w,z)=\rho_{d}(w)H(\rho_{d}(w),z).
\label{FPE3}
\end{eqnarray} 
The solution of Eq. (\ref{FPE2}) with the initial condition 
$\tilde H(w,z)=\delta(w-w(0))$ is  
\begin{eqnarray} &&
\tilde H(w,z)=
\frac{\exp\left\{-[w-w(0)]^{2}/(8D\epsilon^{2}z)\right]}
{(8\pi D\epsilon^{2}z)^{1/2}}.     
 \label{FPE4}
\end{eqnarray}     
Changing variable from $\rho_{d}$ to $p$ while using Eqs. (\ref{FPE3}) 
and (\ref{FPE4}) we obtain 
\begin{eqnarray}
\!F_{d}(p)\!\!=\!\!\frac{\exp\left\{\!-\ln^{2}\!\left[
\mbox{arctanh}\left(p/p_{m}\right)/\rho_{d}(0)\right]/
(8D\epsilon^{2}z)\right\}} {(8\pi
D\epsilon^{2}z)^{1/2}p_{m}\left(1\!-\!p^{2}/p_{m}^{2}\right)
\mbox{arctanh}\left(p/p_{m}\right)} 
 \nonumber \\ &&
 \label{FPE5}
\end{eqnarray} 
for $0<p<p_{m}$ and $F_{d}(p)=0$ elsewhere. Equation (\ref{FPE5}) 
is the same as Eq. (\ref{SSP9}) in section \ref{DCQNLS}.        
A similar calculation based on the Fokker-Planck approach leads 
to Eq. (\ref{div8}) for the $\eta$-PDF for the CQNLSE with disorder 
in the cubic gain/loss coefficient.

\section{Calculation of $F_{d}(p)$ in It\^o's interpretation}
\label{Fd_Ito}
Consider the DCQNLSE with disorder in the linear gain/loss coefficient. 
We now obtain the PDF of $p$ by employing It\^o's interpretation to 
equation (\ref{SSP7}), and show that this PDF exhibits loglognormal 
divergence in the vicinity of $p_{m}$. 
The solution of the equivalent equation (\ref{Ito1}) 
in It\^o's interpretation is \cite{Gardiner2004}  
\begin{eqnarray}&&
\rho_{d}(z)=\rho_{d}(0)\exp\left[2\epsilon X(z)-2\epsilon^{2}z\right],  
\label{Ito2}
\end{eqnarray}   
where the PDF of $X(z)$ is given by Eq. (\ref{div7}). 
Therefore, $X(z)$ is related to $p(z)$ via  
\begin{eqnarray}&&
\!\!\!\!\!\!\!\!\!\!\!\!\!\!\!\!X(z)=\frac{1}{\epsilon}
\left\{\ln\left[
\mbox{arctanh}\left[\frac{p(z)}{p_{m}}\right]/\rho_{d}(0)\right]
+2\epsilon^{2}z\right\}. 
 \label{Ito3}
\end{eqnarray}
Changing variables from $X(z)$ to $p(z)$ we obtain the PDF of 
$p$ in It\^o's interpretation: 
\begin{eqnarray}&&
\!\!\!\!F_{d}^{(I)}(p)\!\!=\!\!
\left[(8\pi D\epsilon^{2}z)^{1/2}p_{m}\left(1\!-\!p^{2}/p_{m}^{2}\right)
\mbox{arctanh}\left(p/p_{m}\right)\right]^{-1}
 \nonumber \\&&
\!\!\!\!\!\times
\exp\left\{\!-\left[
\ln\!\left[\mbox{arctanh}\left(p/p_{m}\right)/\rho_{d}(0)\right]
+2\epsilon^{2}z\right]^{2}/
(8D\epsilon^{2}z)
\right\}.
 \nonumber \\&&
 \label{Ito4}
\end{eqnarray}   
In the vicinity of $p_{m}$ Eq. (\ref{Ito3}) can be approximated by  
\begin{eqnarray}&&
\!\!\!\!\!\!\!\!\!\!\!\!\!\!\!\!
X(z)\simeq\frac{1}{\epsilon}
\left\{\ln\left[
-\frac{1}{2\rho_{d}(0)}\ln\left[\frac{\delta p(z)}{2 p_{m}}\right]\right]
+2\epsilon^{2}z\right\}. 
 \label{Ito5}
\end{eqnarray} 
Consequently, $F_{d}^{(I)}(p)$ is given by 
\begin{eqnarray}&&
\!\!\!\!\left.F_{d}^{(I)}(p)\right|_{p\lesssim p_{m}}\!
\simeq\!\! \left\{(8\pi D\epsilon^{2}z)^{1/2}\delta p
\left|\ln\left[\delta p/(2p_{m})\right]\right|\right\}^{-1}
\times
 \nonumber \\ &&
\!\!\!\!\!\!
\exp\left\{\!-
\left[\ln\!\left[
-\ln\left[\delta p/(2p_{m})\right]/(2\rho_{d}(0))\right]
+2\epsilon^{2}z\right]^{2}
/(8D\epsilon^{2}z)\right\},
 \nonumber \\&&
 \label{Ito6}
\end{eqnarray} 
which exhibits loglognormal divergence as $p$ approaches $p_{m}$.  
We therefore conclude that for the DCQNLSE with disorder in the linear 
gain coefficient both Stratonovich's interpretation and 
It\^o's interpretation lead to loglognormal divergence 
of the PDF of $p$.

\end{document}